\documentstyle[preprint,aps]{revtex}
\begin{document}
\title{Path Integral Density Functional Theory}
\author{Peter Borrmann}
\address{Department of Physics, Carl von Ossietzky
Universit{\"a}t Oldenburg, D-26111 Oldenburg, Germany}
\maketitle
\narrowtext
\begin{abstract}
A new method ( PI-DFT ) which combines path integrals and
density functional theory is proposed as a  pathway to many
fields of physics. Within path integral theory it is possible to
construct particle densities without explicitly calculating
individual wave functions.
These densities can directly be used as
an input to energy density functionals.
Thus our method makes full use of the theorem of
Hohenberg, Kohn and Sham which shows, that the energy of a many electron system
only depends on the particle density.\\
At glance we present
a recursion formula for the calculation of many fermion and boson
particle densities from  one-particle densities at a set of different
temperatures.
For both statistics the numerical effort of our method increases
only with the square of the particle number.
\end{abstract}
%
%
Path integrals and Density Functional Theory have both been investigated
and developed  extensively in the past decades
\footnote{See for example \cite{jones,klein1} and references therein.}.
Here we present a new method (PI-DFT),
which formally joins both fields and for the first time
uses the theorem of Hohenberg, Kohn
and Sham \cite{kohn1,kohn2} to its full
extent. While the method is applicable to bosons
as well, the most interesting case is that of fermions
because of its great relevance in quantum chemistry. The
pride of our method is that we never deal with individual wave functions
to compute the particle densities but still solve the full problem.
The basic underlying idea is a self-consistent
iteration scheme, in which the
particle density is computed by means of path integrals, while effective
potentials and observables are computed with methods lent from density
functional theory. \\
For the problem of constructing
completely antisymmetric
or symmetric particle densities we propose a recursion formula
which  has originally been invented by us for the partition
function \cite{borr1}. With this formula there is no need to know
the individual wave functions to construct the appropiate
particle densities for Fermi-Dirac (FD) or Bose-Einstein (BE)
statistics. Instead,  these are
calculated from single-particle densities but at different temperatures.\\
Using this formula the numerical effort for both statistics grows
only with the square of the particle number.\\

In the following we will consider N-particle hamiltonians of the from
\begin{equation}
H = - \frac{\hbar^{2}}{2m}\sum_{i=1}^{N} \frac{{\rm d}^{2}}{{\rm d}{\vec
x}^{2}}
+ \sum_{i=1}^{N} v(\vec{x}_{i})
+ \sum_{k<l}^{N} u(\mid \vec{x}_{k} - \vec{x}_{l} \mid) \; .
\end{equation}
That is, we assume the potential to contain one and two particle interaction
terms
as encountered typically in quantum chemical problems.

In conventional Path Integral Monte Carlo techniques for fermion systems
an $N M$ dimensional integral
 with an integrand containing an ($N \times N$) determinant
has to be calculated ( see e.g. \cite{ti1,ti2})
resulting  in a numerical
effort proportional to $N^{4}$.
Equation (\ref{trotpath}) can be derived easily using the Trotter formula
\cite{trott}
and is equivalent to the thermal Path Integral as the number of {\sl time
slices} $M$
goes to
infinity.
\begin{eqnarray}
\label{trotpath}
Z&=& \left( \frac{1}{N!} \right)^{M} \int \left[ \prod_{k=1}^{N}
\prod_{\mu=1}^{M}
{\rm d}\vec{x}_{k}(\mu) \right]  \\
&& \nonumber
\prod_{\nu=1}^{M} \left\{ \det A(\nu)
\times \exp\left( - \frac{\beta}{M} \left[ \sum_{k=1}^{N} v(\vec{x}_{k}(\nu))
+ \sum_{k<l}^{N} u( \mid \vec{x}_{k}(\nu)-\vec{x}_{l}(\nu) \mid ) \right]
\right) \right\}
\end{eqnarray}
with
$$
(A(\mu))_{k,l}=\left( \frac{M m}{2 \pi \beta \hbar^{2}} \right)^{3/2}
\exp\left( -\frac{M m}{2\beta\hbar^{2}} (\vec{x}_{k}(\mu) -
\vec{x}_{l}(\mu+1))^{2}
\right) \; ,
$$
$$
\vec{x}_{k}(M+1)\equiv \vec{x}_{k}(1) \; .
$$
Although the method is suitable for systems like $^{3}$He clusters \cite{bh92}
and most physical relevant observables can be calculated, its application is
numerically very extensive in general. \\

According to the theorem of Hohenberg, Kohn and Sham  \cite{kohn1,kohn2} and
its  extension to thermal systems given by Mermin \cite{mermin} the potential
given by the term in square brackets in (\ref{trotpath})
can be replaced by an effective
one-particle potential.
\begin{equation}
\label{effpot}
\varphi(\vec{x}) = v(\vec{x}) + \int {\rm d}\vec{x'}
\eta(\vec{x'}) \; u( \mid \vec{x'} - \vec{x} \mid ) \; .
\end{equation}
Since the energy is only a functional of the one particle density
$\eta(\vec{x})$,
which can be easily calculated  within the approximation
(\ref{trotpath}) by Monte Carlo methods
and  $\varphi(\vec{x'})$, the solution
of the stated problem, can be found by
self-consistent iteration of (\ref{trotpath})
 and
(\ref{effpot}). A priori the use of this
approximation has no advantages compared to the
direct calculation of (\ref{trotpath}), but
together with a much more effective way of calculating
the partition function, which will
be stated in the following, the above iteration
scheme develops to a very powerful tool.

Let us denote with  $\eta_{N}(\vec{x};\beta)$ the probability of finding
a particle at position $\vec{x}$ in a system
of N noninteracting  fermions or bosons moving in a common potential
thermalized at temperature $\beta$ and with $Z_{N}(\beta)$ the
corresponding canonical partition function.
For such a system it is sufficient
to compute the one-particle densities $\eta_{1}(\vec{x};\beta),
\eta_{1}(\vec{x};2 \beta), \ldots,  \eta_{1}(\vec{x};N \beta)$,
but at a set of different temperatures.  Then
$Z_{N}(\beta)$ and $\eta_{N}(\vec{x},\beta)$
can be calculated by the recursion formulas
\begin{equation} \label{zrecu}
Z_{N}(\beta) = \frac{1}{N} \sum_{k=1}^{N}  (\pm)^{k+1} Z_{1}(k \beta)
\; Z_{N-k}(\beta)
\end{equation}
\begin{equation} \label{grecu}
\eta_{N}(\vec{x};\beta) = Z_{N}^{-1}(\beta)  \sum_{k=1}^{N} (\pm)^{k+1}
Z_{1}(k \beta) \;  \eta_{1}(\vec{x}, k\beta) \; Z_{N-k}(\beta)
\end{equation}
where the plus and the minus signs stand for Fermi-Dirac and
Bose-Einstein statistics, respectively, and $Z_{0}(\beta) \equiv 1$.\\
The proof for (\ref{zrecu}) has already
been given explicitly in \cite{borr1}.
The validity of (\ref{grecu}) can be proven in the same manner.  As
the whole procedure is somewhat lengthy, we omit it here and sketch
the basic idea using the two-fermion system as an example.

Let $\epsilon_{k}$ denote the energy eigenvalues of the one-particle
system and $\psi_{k}$ the corresponding eigenfunctions.
Then the partition function is given  by
\begin{equation}
Z_{2}(\beta)= \frac{1}{2} \sum_{k} \sum_{l \neq k}
\exp\left(-\beta ( \epsilon_{k}+\epsilon_{l}) \right) \; .
\end{equation}
The probability $p_{k}$ of finding a
particle in the one-particle state
$\mid k \rangle$ is simply
\begin{eqnarray}
p_{k}&=&  Z_{2}^{-1}(\beta) \;  \frac{1}{2} \;
 \sum_{l \neq k} \exp(-\beta (\epsilon_{k}+\epsilon_{l}))\\
&=&   Z_{2}^{-1}(\beta) \; \frac{1}{2} \;
 \sum_{l} \exp \left( - \beta ( \epsilon_{k} + \epsilon_{l} ) \right)
- \exp \left( - 2 \beta \epsilon_{k}  \right)   \nonumber \\
 &=&  Z_{2}^{-1}(\beta) \;  \frac{1}{2} \;
 \left[ \exp( -\beta \epsilon_{k} ) \;
 Z_{1}(\beta) - \exp(-2 \beta \epsilon_{k} ) \right]   \nonumber
\end{eqnarray}
For the particle density this yields
\begin{eqnarray}  \label{twodens}
&&\eta_{2}(\vec{x};\beta)= \sum_{k} p_{k}
 \mid \psi_{k}(\vec{x}) \mid^{2} \\
&&= Z_{2}^{-1}(\beta) \; \sum_{k} \mid \psi_{k} (\vec{x}) \mid^{2}
 \left[ \exp(-\beta \epsilon_{k}) \; Z_{1}(\beta) -
\exp( -2 \beta \epsilon_{i} ) \right]       \nonumber  \\
&&= Z_{2}^{-1}(\beta) \left[ Z_{1}^{2}(\beta) \; \eta_{1}(\vec{x};\beta)
- Z_{1}(2 \beta) \; \eta_{1}(\vec{x};2 \beta) \right] \; , \nonumber
\end{eqnarray}
which is nothing but equation (\ref{grecu}) for
the special case  $N=2$. \\
A simple example shows that the given formulas work very well and
exactly. We calculated the exact solvable system of $N$ uncoupled harmonic
oscillators.
\begin{equation}
H = - \frac{\hbar^{2}}{2m}
\sum_{k}^{N} \frac{{\rm d}^{2}}{{\rm d}x_{k}^{2}}
+ \frac{1}{2} x_{k}^{2}
\end{equation}
for both statistics. The one particle densities
$\eta_{1}(\vec{x};k \beta)$
shown in Fig.~1  are calculated
 with standard path integral techniques
(see e.g. \cite{ti1,ti2}). In Fig.~2 and Fig.~3
the exact particle densities
and the densities calculated via (\ref{grecu}) are shown for both
statistics. The outcome is an almost perfect agreement with the
exact results. The computed energies agree equally well with the
exact values. For our examples the errors were in all cases below
0.1~\%.\\
Obviously  it is essential in the fermion case, that there is
a nonvanishing occupation probability of the excited states for the
chosen temperature, i.e. the results of the one-particle Path
Integral Monte Carlo simulations have to contain information about
the excited states. Because of the fact that in the low temperature
range the energy is in  general
a very slowly increasing function in the case of Fermi-Dirac statistics,
this is not a too bad restriction, even if one
is going to calculate ground state properties of the fermion system.\\

Possibly the most interesting topic concerning PI-DFT is an analysis of
its computational costs.
The basic steps in PI-DFT are the following:
\begin{enumerate}
\item Construction of an initial guess for the N particle density.
\item Calculation of the effective potential.
\item Calculation of single-particle densities
      and the corresponding energies at N different temperatures.
\item Calculation of the partition function by spline interpolation
      of the caloric curve along with numerical integration using
      \begin{equation}
       Z(\beta) ) = \exp( - \int_{\beta_{0}}^{\beta}
       {\rm d}\beta' E(\beta')  + F_{0})
       \; . \nonumber
      \end{equation}
\item Calculation of the N-particle partition function using (\ref{zrecu}).
\item Calculation of the N-particle density using
      (\ref{grecu}).
\end{enumerate}
Steps 2-6 have to be repeated until the particle densities and the
energy are converged. The main numerical effort lies in the steps
2 and 3. The calculation of the effective potential is usually
proportional to $N^{2}$. Because of
the fact that the numerical effort of a Path Integral Monte Carlo
simulation grows normally proportionally to $\beta$ the same applies to
step 3.  Steps 4 and 5 are negligible, while the effort
for step 6 again grows proportionally to $N^{2}$ assuming that
the number of used grid points is approximately proportional to
the number of particles.
Overall this indeed yields a numerical effort proportional to the
square of the particle number  for PI-DFT. Hence it follows
that PI-DFT can be seen
as a formal proof, that e.g. quantum chemical calculations can be
done with such a slowly increasing numerical effort. We hope that
this will encourage  further improvements of other related methods, too.
This paper presents only the onset of PI-DFT and a lot of
improvements seem to be possible and required. For example, the
recursion for the partition function might be numerically unstable
under unfavorable circumstances in the case of FD statistics because
of the alternating sums.\\
In a forthcoming paper we will present a study on $^{4}$He clusters
as a first physical application of PI-DFT.
\acknowledgments
We wish to thank E.R. Hilf for very fruitful and
helpful discussions.
%
%

%
%
\begin{figure}
\label{pathdens}
\caption{Particle densities $\eta_{1}(\vec{x};\beta)$ calculated with
Path Integral Monte Carlo for various temperatures ($\hbar=k=m=1$).}
\end{figure}
\begin{figure}
\label{fdtdens}
\caption{Exact harmonic oscillator particle densities and particle
densities calculated  with Path Integral Monte Carlo in connection with
equation (3) for N=2 at T=0.5 and N=4 at T=2.0 for the case of FD
statistics ($\hbar=k=m=1$).}
\end{figure}
\begin{figure}
\label{bodens}
\caption{Exact harmonic oscillator particle densities and particle
densities calculated  with Path Integral Mont Carlo in connection with
equation (3) for n=2 at T=0.5 and N=4 at T=1.0 for the case of BE
statistics ($\hbar=k=m=1$).}
\end{figure}
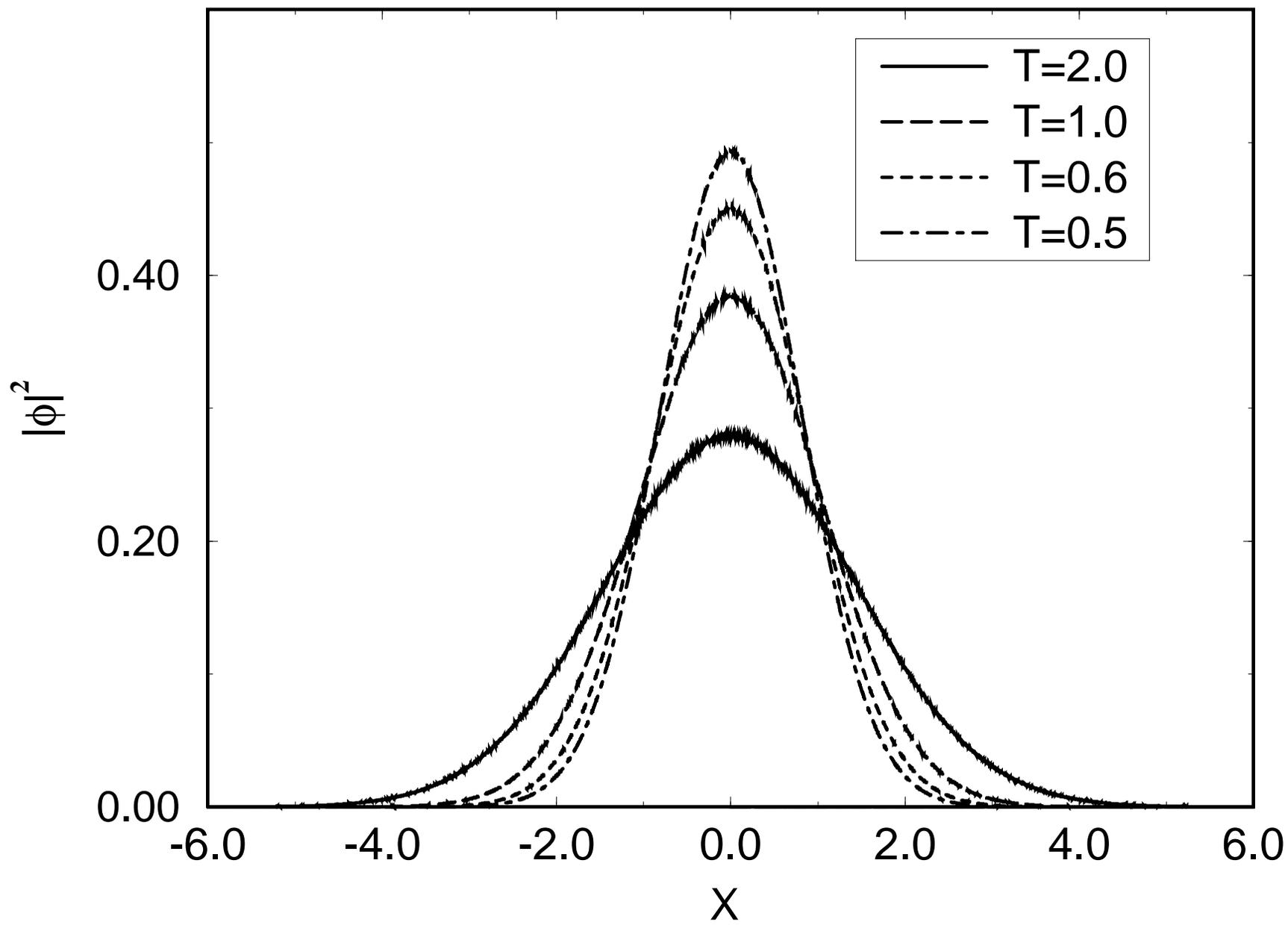
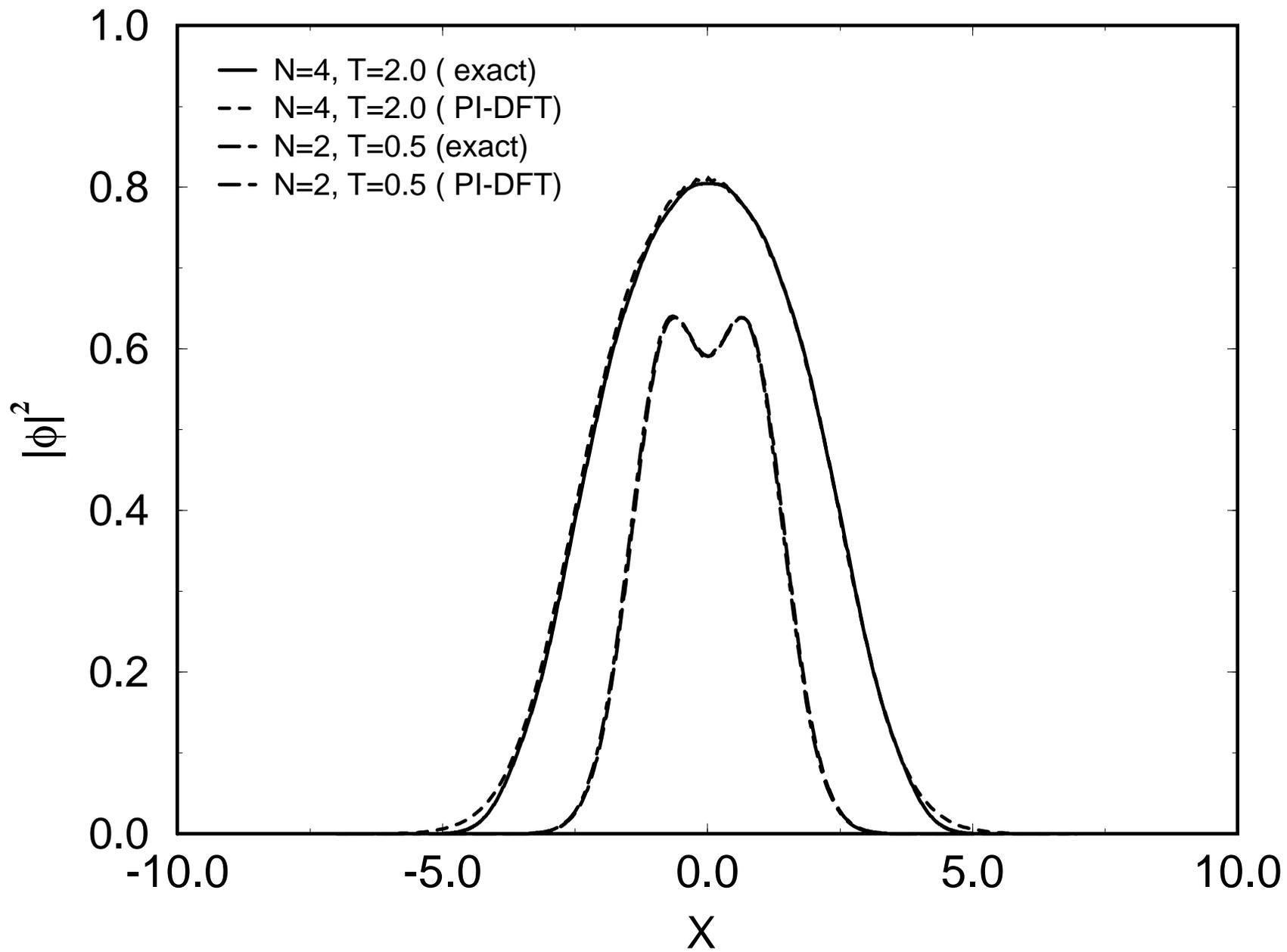
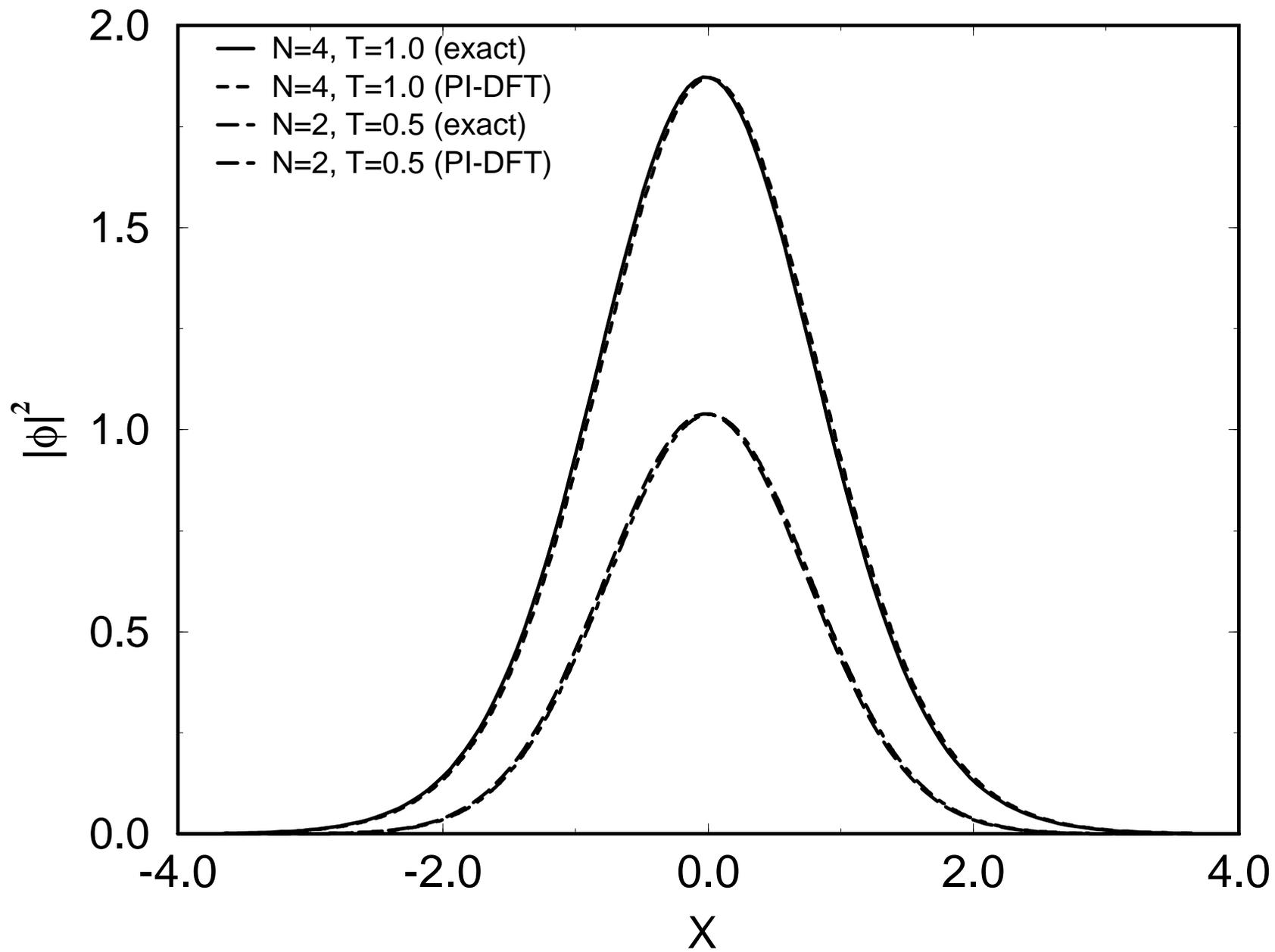
\end{document}